\pdfoutput=1                

\documentclass[12pt,letterpaper]{article}

\usepackage[utf8]{inputenc}
\usepackage[T1]{fontenc}
\usepackage{lmodern}            
\usepackage[affil-it]{authblk} 
\usepackage{enumitem}
\usepackage{float}
\usepackage{amsmath}          

\usepackage{geometry}
\usepackage{setspace}
\usepackage{amsmath,amssymb,amsfonts,amsthm}

\usepackage{graphicx}
\usepackage[table,dvipsnames]{xcolor}

\usepackage{booktabs,longtable,multirow,array}

\newcolumntype{C}[1]{>{\centering\arraybackslash}m{#1}}

\usepackage[colorlinks=true,
            linkcolor=BrickRed,
            citecolor=SeaGreen,
            urlcolor=MidnightBlue]{hyperref}

\usepackage[font=footnotesize,labelfont=bf,labelsep=period]{caption}

\usepackage[numbers,sort&compress]{natbib}
\definecolor{bloodorange}{HTML}{D15500}
\definecolor{darkgreen}{HTML}{004d00}

\title{
    \vspace{-2cm} 
    {\fontsize{20pt}{26pt}\selectfont \textbf{\textcolor{bloodorange}{The Attribution Crisis in LLM Search Results}}}\\[-0.5mm]
    {\Large {{Estimating Ecosystem Exploitation}}}\\[1.cm]
}
\date{}

\author[1,2]{Ilan Strauss}
\author[4]{Jangho Yang}
\author[1,3]{Tim O'Reilly}
\author[1]{Sruly Rosenblat}
\author[1]{Isobel Moure \thanks{We gratefully acknowledge funding support from The Alfred P. Sloan Foundation, the Omidyar Network, and the Patrick J. McGovern Foundation. We extend our appreciation to the people we have had conversations with that have shaped our thinking. \emph{Contact:} \href{mailto:istrauss@ssrc.org}{istrauss@ssrc.org / ilan@aidisclosures.org}. \emph{Code and Data:} \href{https://github.com/AI-Disclosures-Project/Ecosystem_Exploitation_In_Search_Results}{https://github.com/AI-Disclosures-Project/Ecosystem\_Exploitation\_In\_Search\_Results}.}}

\affil[1]{AI Disclosures Project (Social Science Research Council)}
\affil[2]{Institute for Innovation and Public Purpose (University College London)}
\affil[3]{O'Reilly Media}
\affil[4]{University of Waterloo}
\begin{document}

\maketitle
\begin{abstract}\onehalfspacing

Web-enabled LLMs frequently answer queries without crediting the web pages they consume, creating an ``\textit{attribution gap}'' -- the difference between \textit{relevant} URLs read and those actually cited. Drawing on approximately 14,000 real-world LMArena conversation logs with search-enabled LLM systems, we document three exploitation patterns: 1) \textbf{No Search}: 34\% of Google Gemini and 24\% of OpenAI GPT-4o responses are generated without explicitly fetching any online content; 2) \textbf{No citation}: Gemini provides no clickable citation source in 92\% of answers; 3) \textbf{High-volume, low-credit}: Perplexity's Sonar visits approximately 10 relevant pages per query but cites only three to four. A negative binomial hurdle model shows that the average query answered by Gemini or Sonar leaves about 3 relevant websites uncited, whereas GPT-4o's tiny uncited gap is best explained by its selective log disclosures rather than by better attribution. Citation \textit{efficiency} -- extra citations provided per additional relevant web page visited -- varies widely across models, from 0.19 to 0.45 on identical queries, underscoring that retrieval design, not technical limits, shapes ecosystem impact. We recommend a transparent LLM search architecture based on standardized telemetry and full disclosure of search traces and citation logs.


\end{abstract}

\thispagestyle{empty}
\newpage

\tableofcontents
\thispagestyle{empty}
\newpage
\setcounter{page}{1}
\pagenumbering{arabic}

\section{Introduction}
\setcounter{page}{1}
\pagenumbering{arabic}

\begin{quote}
``The misery of being exploited by capitalists is nothing compared to the misery of not being exploited at all.'' -- Joan Robinson, \textit{Economic Philosophy}, 1962.
\end{quote}

The rapid integration of large language models (LLMs) into information-seeking workflows has fundamentally transformed how users access and interact with content on the world wide web. Search-augmented LLMs, which combine generative capabilities with real-time web retrieval, promise to deliver more accurate, up-to-date, and comprehensive responses than traditional chatbots limited to just their training data \citep{huang2023citation}. But questions have been raised about source accuracy \citep{npr_trump_health_report_2025}, intellectual property rights, scraping websites without permission -- so-called `access violations' \citep{lucky_2025, rosenblat2025beyond}, and what we term \textbf{``ecosystem exploitation''}-- \textit{the gap between relevant web content consumed by the LLM when answering a query and those sources cited in the model's output (its response)}.\footnote{We use `consume', `read', and `search' interchangeably for a relevant website visit logged by the LLM.}  

This \textit{attribution gap} has serious implications for the digital ecosystem on which AI's ongoing utility depends. Content creators, publishers, and knowledge producers rely on appropriate attribution and licensing agreements when their information is used to answer queries. When LLMs systematically consume relevant content without adequate citation or remuneration, they undermine the incentive structures that support high-quality information production and threaten the economic viability of content creation at scale \citep{oreilly_ai_original_sin}.\footnote{For previous research empirical research on attribution practices across commercial LLM systems see \citet{gao2023alce, yue2023automatic, li2023survey, huang2025authorship, fayyazi2025lea, tryprofound_citation_patterns_2025}.} This systematic lack of attribution of the content sources consumed by LLMs during web search is a widely known issue \citep{autogpt_perplexity_controversy_2024, hacker_news_claude_citations_2025, reuters2025bbc}.\footnote{As recently as June 2025 the BBC threatened Perplexity's search system with legal action for using its content during search \citep{reuters2025bbc}. Perplexity has set-up a revenue-sharing program with some publishers and increased citation of sources in response to accusations against it \citep{autogpt_perplexity_controversy_2024}.} Yet major LLM vendors reveal little about how their retrieval-augmented generation (RAG) pipelines choose and ingest web content, and how they cite appropriate web sources \citep{stox2025trafficmentions}.

\noindent \textbf{\textsc{Research questions}}: \textit{To what extent do search-augmented LLMs exploit the web — using content without citing it? And what distinguishes today's best and worst attribution practices?}\\[-3mm]


\noindent \textbf{\textsc{Method and Data}}. Using an LMArena dataset of $\approx$ 14,000 real-world answers to around 7,000 multi-turn queries taken between March and April 2025, we analyze the ``attribution gap'' exhibited by 11 LLM search-enabled models across three provider families: OpenAI (GPT-4o), Perplexity (Sonar), and Google (Gemini). We define the attribution gap as the number of \textit{relevant} URLs visited by the LLM system when answering the query minus the number of URLs cited in the model's output, providing a direct measure of ecosystem exploitation. To isolate this quantity, we remove all hallucinated citations (numbered citations provided in-text that do not have a corresponding source in the search log) and all ungrounded citations (URL citations provided that do not appear in the search logs but may link to a valid site). Our statistical model is a negative binomial hurdle model with bootstrapped confidence intervals. This allows us to quantify both when attribution gaps occur and how severe they are, while accounting for differences in query type (e.g. `Data Science', `Current Affairs', etc.). In a second regression we leverage the dataset's head-to-head design, looking at differences between model citation behavior for the same query.\\[-2mm]

\noindent \textbf{\textsc{Key findings}}: 

\begin{enumerate}

\item \textbf{Search-enabled LLM systems exploit by, surprisingly, not searching at all}, relying instead on their pre-training data or simply not disclosing relevant search logs accurately. Despite the models being in search mode, 15.6\% of LLM answers skipped web search entirely. This was highest for Google's Gemini (34\%), followed by OpenAI's GPT-4o models (24\%). 

\item \textbf{LLM search systems exploit by providing no citations (zero attribution)}. 30\% of answers provided no citations. This is driven less by query topic and more by model-specific behaviors. Gemini provided no citations for a striking 92\% of queries, undermining claims that its impact on third-party traffic will be negligible. 
  
\item Our zero-hurdle statistical model shows that, for a typical query, \textbf{Google's Gemini models and Perplexity's Sonar models have sizable attribution gaps}: at approximately 3 relevant websites visited but not cited. Perplexity exhibits much higher volume ecosystem exploitation, visiting $\approx$10 relevant websites per query, but with a similar overall attribution gap to Gemini.

\item \textbf{The full extent of ecosystem exploitation may be underestimated} because models appear to selectively disclose which websites they visit, especially GPT-4o models. By design, GPT-4o models appear to have a near perfect correspondence between relevant websites drawn on and those cited, leading to a small attribution gap.

\item \textbf{Refining a search-LLM's RAG pipeline can almost double the number of citations it provides for each extra webpage it consumes}. In ``head-to-head'' model regressions, comparing citation differences between model pairs for identical queries, we find that citation efficiency -- the extra citations shown per additional website visited -- ranges from 0.19 to 0.45. This indicates that retrieval design (reasoning modules, search context size, and geolocation), rather than technical limits, determines AI's relationship with the world wide web's ecosystem.
        
\end{enumerate}

The classical political economists defined \textit{exploitation} as a category of production, whereby an owner-producer appropriates the difference between the cost of an input and its realized value contribution to output \citep{sep-exploitation}. The classical economists focused exclusively on the labor input \citep{fine1989marx, hollander1992classical}, but we can easily extend this framework to the data inputs consumed by LLM models during inference when producing a relevant response (the output).

\noindent\textbf{\textsc{Policy implications}}.  
Advancing a healthy web ecosystem requires \emph{transparent search telemetry} (logs, traces, and metrics). Developers, enterprise buyers, and potentially regulators should insist that LLM APIs expose a standard trace of every retrieval step and the sources ultimately cited.  The tooling already exists to implement this: observability stacks such as LangSmith, Langfuse, Phoenix, and the GenAI semantic-conventions in OpenTelemetry can record an end-to-end \emph{search trace} — query, retrieval, re-ranking, and citation — so long as each web document is tagged with a stable source-ID (typically a URL hash).\footnote{See: \url{https://opentelemetry.io/docs/specs/semconv/gen-ai/}.} Per-document relevance scores can travel in the same span \citep{langchain_rag_citations}. If all providers adopt common definitions (e.g.\ \texttt{llm.retrieval.ids}, \texttt{llm.retrieval.scores}), third-parties could compare across models the exact ratio of ``information consumed'' to ``information cited''; and equitable business models could be built on top of this.

\noindent \textbf{Limitations}. Our study does not explicitly test citation accuracy or relevance, beyond removing hallucinated and ungrounded citations \citep{gao2023alce}. Additionally, our study does not account for access violations \citep{lucky_2025, rosenblat2025beyond} -- whether the LLM had permission to visit specific websites, which may be governed by licensing agreements. Despite these limitations, our study represents the first systematic, cross-model audit of attribution behavior in commercial search-augmented LLM systems, focusing specifically on their search tools. Our goal is to provide a structured framework for assessing attribution in empirical LLM studies \citep{bbc2025research}. 

Section \ref{sec:methods} describes our data and variables; Section \ref{sec:desc} details key empirical features of our data; Section \ref{sec:regmodels} outlines our two regression models; Section \ref{sec:findings} presents our core findings from the regressions; Section \ref{sec:discussion} discusses policy implications; and Section \ref{sec:conclusion} concludes. Our Appendix \ref{sec:appendix} contains more detailed model results.

\newpage

\section{Data and Method}\label{sec:methods}

We conduct a large-scale empirical audit of attribution practices of commercial search-augmented LLMs in real-world user interactions. \textit{Attribution} refers to identifying the source material or input features that contributed to a model's generated output or decisions \citep{li2023survey}. It emphasizes \textit{relevant} content ``consumed'' (read and used) by the LLM when answering, instead of all websites visited that may not be relevant and therefore not drawn upon. We assume the search results provided in an LLM's search log were already determined to be relevant by the provider. Other forms of exploitation, especially ones involving unauthorized access are not examined here \citep{lucky_2025, oreilly_ai_original_sin}. \\[-4mm] 

\noindent \textbf{\textsc{Dataset Overview}}. Our analysis uses LMArena's search dataset, containing pairs of model answers to the same user query across 11 major commercial LLMs. Our final sample size (n) is 13,929 observations. We construct all variables from the raw logs rather than using the variables provided by LMArena due to various errors and inconsistencies.\footnote{See: \url{https://blog.lmarena.ai/blog/2025/search-arena/}.}  

The initial dataset before filtering contains 14,000 conversations from 3,642 users, covering 7,000 queries each given to a pair of models between March (44\%) and April (56\%) 2025. Each query represents a potential multi-turn conversation, including the model's final response. Crucially, this dataset captures actual deployment behavior via application programming interface (API) calls.\\[-4mm]

\textbf{Model Coverage}. We analyze 11 commercial variants grouped by provider:
\begin{itemize}
\item \textit{OpenAI}: api-gpt-4o-mini-search, api-gpt-4o-search, api-gpt-4o-search-high, api-gpt-4o-search-high-loc
\item \textit{Perplexity}: ppl-sonar-pro, ppl-sonar-reasoning, ppl-sonar, ppl-sonar-pro-high, ppl-sonar-reasoning-pro-high  
\item \textit{Google}: gemini-2.0-flash-grounding, gemini-2.5-pro-grounding
\end{itemize}

The model configurations are detailed further in the LMArena documentation.\footnote{``For Perplexity and OpenAI, this includes setting the `search context size' parameter to medium, which controls how much web content is retrieved and passed to the model. We also explore specific features by changing the default settings: (1) For OpenAI, we test their geolocation feature in one model variant by passing a country code extracted from the user's IP address. (2) For Perplexity and OpenAI, we include variants with `search context size' set to high. '' Gemini model defaults to Google Search Tool enabled. See: \url{https://blog.lmarena.ai/blog/2025/search-arena/}. Accessed: 16 June, 2025.} 

\subsection{Key Variables and Measurement}

We do not rely on LMArena's citation variable construction, which we found didn't adequately distinguish between search logs and citations for our use case. We instead reconstruct them from scratch using the model logs in their dataset.\\[-4mm]

\textbf{Ecosystem Exploitation (Attribution Gap)}. Our primary dependent variable measures the difference between unique relevant pages visited by the LLM during search and unique pages actually cited to the user in the API response. This captures the extent to which a model consumes relevant web content without providing appropriate attribution.

\textbf{Citations}. We define citations as any in-text URL reference that is grounded in the model's search results log, including both explicit URLs and numbered references that link to specific sources. We exclude ``hallucinated citations'' (numbered citations, such as $^{[13]}$, that link to an empty source) and ``ungrounded citations'' (citation of URLs that exist but are not in the search log). Unique URLs (websites) for citations or attribution are those seen by the model that do not link to the same web page when stripped of parameters (e.g., \url{example.com/?tracking_id=23222} is turned into \url{example.com}). This amounts to transforming URLs into their base URL and then checking for duplicates.


\textbf{Relevant Sites Visited (Consumed)}. We define relevant sites visited by the LLM as those listed in their search log, regardless of whether they were cited in-text or not. We treat every URL that appears in the search log as relevant, on the understanding that the provider has already filtered out non-relevant visits. Note, however, that this log may itself be incomplete: some vendors — OpenAI, in particular — seem to pre-trim their traces, returning only a subset of the relevant pages the model actually visited.

\textbf{Conversation Classification}. We categorize all conversations into 12 topic areas using o4-mini with high reasoning to enable analysis of topic-specific attribution patterns. This classification covers areas from technical queries (software engineering, data science) to consumer-oriented topics (shopping, health, finance). We provided the 12 categories to GPT, drawing on an unsupervised visualization of clusters in the query data on Nomic Atlas.\footnote{See Nomic Atlas: \url{https://atlas.nomic.ai/data/srulyrosenblat/ai-model-search-comparison-dataset/map/2da8af3f-c160-4008-928d-8df37c27b947\#x7Nh}.}

\textbf{Data Cleaning}. After removing conversations with classification failures or misaligned search traces, our final sample contains 13,929 observations, all successfully classified and with clean attribution data.

\textbf{Pre-filtering of Logs}. Most models appear to filter their logs for relevant websites visited only -- rather than all sites visited. But OpenAI appears to filter them more strictly. Hallucinated citations (numbered citations provided in-text that do not exist in the search log) appear for Gemini and to a lesser extent Sonar.\footnote{They are by definition zero for GPT-4o models since OpenAI's models do not use numbers in square brackets for its citations which is what we track for this metric.} Moreover, in terms of \textit{citations provided that do not appear in the search logs} (ungrounded citations): 7\% of  citations by GPT-4o models were not found in its search logs, indicating either an overly restrictive pre-filtering of the search logs or simply the model citing from pre-trained knowledge.

\section{Descriptive Patterns} \label{sec:desc}
Our data show stark differences in attribution practices across model families, highlighting that design choices, rather than technological limitations, drive model behavior.\\[-4mm]

\noindent \textbf{\textsc{The Attribution Crisis}}. 15.6\% of LLM responses involved no website visits despite being in search mode, yet 30\% provided no citations whatsoever -- a substantial gap between content consumption and content recognition. This pattern varies dramatically by provider. 39\% of responses show perfect attribution alignment (zero gap), while 61\% exhibit some degree of ecosystem exploitation. For Gemini, a ``zero'' gap still tends to signals exploitation though, because the model usually answers without visiting any external site even when its Google-Search tool is enabled (Table \ref{tab:model_family_summary}).\\[-5mm]

\noindent \textbf{\textsc{Provider-Specific Patterns}}. Table~\ref{tab:model_family_summary} illustrates three distinct patterns of exploitation: (1) \textbf{Not visiting relevant sites} at all, according to the model's logs, when answering a question (exploitation through pre-training reliance) -- 24\% of GPT search answers and 34\% of Gemini models; (2) Not providing \textbf{any citations} at all -- 25\% of GPT-4o answers and 92\% of Gemini answers; and (3) Having a \textbf{large relative gap} between sites visited and sites cited -- as in Perplexity's Sonar.\\[-3mm]

\begin{table}[H]
\centering
\rowcolors{2}{gray!15}{white}
\caption{\centering {\large{Attribution Statistics by Model Family}}}
\vspace{-2mm}
\begin{tabular}{lrrr}
  \toprule
  & GPT-4o & Gemini & Sonar \\ 
  \midrule
  Median Gap           & 0.0     & 4.0     & 5.0     \\ 
  Median Citations     & 2.0     & 0.0     & 5.0     \\ 
  Median Sites Visited & 2.0     & 4.0     & 10.0    \\ 
  Zero Citations       & 1,309   & 2,198   & 663     \\ 
  Zero Site Visits     & 1,242   & 817     & 120     \\ 
  Sample Size (n)      & 5,272   & 2,394   & 6,263   \\
  \bottomrule
\end{tabular}
\vspace{2mm}
\caption*{Note: `Gap' refers to the attribution gap = relevant websites visited (content consumed) minus unique websites cited.} 
\vspace{-4mm}
\label{tab:model_family_summary}
\end{table}

\begin{itemize}
\item \textbf{GPT-4o -- Limited disclosure of sites visited}: On paper it shows almost perfect alignment between relevant pages searched according to its logs and pages cited. But this is likely an artifact of aggressive log-filtering. The model seems to disclose only those URLs it ultimately cites, omitting any additional relevant pages it read (consumed). Support for this view comes from its high share of \emph{ungrounded} citations — links that appear in the answer but not in the trace — suggesting that many visits are simply withheld from the log.

\item \textbf{Gemini -- No citations provided}: Systematic attribution failure with 92\% zero-citation responses despite some relevant website searches.

\item \textbf{Sonar -- High-volume, low-credit}: Extensive crawling (10 median sites) with large attribution gaps (5.0 median) despite providing more citations than Gemini or OpenAI.
\end{itemize}


\noindent \textbf{\textsc{Topic-Specific Vulnerabilities}}. Attribution failures concentrate in economically and legally sensitive domains, including:

\begin{itemize}
\item \textit{Software engineering}: 33\% zero-citation rate (700 queries), especially for Gemini
\item \textit{Games/Creative writing}: 40\% zero-citation rate (494 queries) 
\item \textit{Education}: 43.6\% zero-citation rate (228 queries) 
\item \textit{Health information}: For Mental \& Physical Health and Relationships, systematic gaps for Gemini (94\%) and GPT (26\%)
\end{itemize}

\section{Regression Models} \label{sec:regmodels}
\subsection{Hurdle-Model Specification}

The dependent variable \(Y_i\;(i=1,\dots,n)\) is the \emph{attribution gap}: the number of relevant websites a model visits but fails to cite, for query \(i\), where $i$ runs from $1$ to $n=13,929$. Because most answers are perfectly attributed (\(Y_i=0\)) while the rest show a skewed count distribution, we use a \textbf{hurdle model}. It breaks up $E[Attribution \ \ Gap] = P(Gap > 0) \ \times \ E[Gap \ | \ Gap > 0]$ into two sequential components:

\begin{enumerate}[label=(\alph*),itemsep=2pt,leftmargin=1.4em]
\item \emph{Hurdle stage:} What is the probability that the gap is exactly zero $\rightarrow$ $P(Gap > 0)$? This is a Bernoulli outcome with probability \(\pi_i\).
\item \emph{Count stage:} If an attribution gap exists (\(Y_i>0\)), how large is the gap: $E[Gap \ | \ Gap > 0]$? We model positive counts with a \emph{zero-truncated} negative binomial distribution, which has mean \(\lambda_i\) and over-dispersion parameter \(\alpha\).\footnote{If \(\alpha\to0\) the model would reduce to a Poisson hurdle.}
\end{enumerate}

Putting the two parts together gives the probability mass function:  

\[
\Pr(Y_i=y \mid \mathbf{x}_i)=
\begin{cases}
\pi_i, &
y=0,\\[6pt]
(1-\pi_i)\,
\dfrac{f_{\mathrm{nb}}\!\bigl(y;\lambda_i,\alpha\bigr)}
      {1-f_{\mathrm{nb}}\!\bigl(0;\lambda_i,\alpha\bigr)}, &
y\ge 1,
\end{cases}
\tag{1}\label{eq:pmf}
\]

\noindent where:  

\[
f_{\mathrm{nb}}\!\bigl(y;\lambda,\alpha\bigr)=
\frac{\Gamma\!\bigl(y+\alpha^{-1}\bigr)}
     {\Gamma(\alpha^{-1})\,y!}\,
(\alpha\lambda)^{y}\,
\bigl(1+\alpha\lambda\bigr)^{-\bigl(y+\alpha^{-1}\bigr)},
\qquad \alpha>0 .
\tag{2}\label{eq:nb-pmf}
\]

\noindent \(f_{\text{nb}}(y;\lambda,\alpha)\) is the mean-parameterized negative binomial probability mass function,

\medskip
\noindent\textbf{Linear predictors.} Each of the two regression stages has its own regression equation with predictors, consisting of: model family, the query category or `classification' (e.g. `Data Science'), the log of LLM output character count, and the number of unique search results. This leads to the following two equations:\footnote{Number of `turns' had poor predictive power in our model, wrong sign in count component, despite some correlation to attribution gap for GPT models and especially for Perplexity models.}
\vspace{1mm}
\begin{align}
\text{logit}(\pi_i) &=
  \gamma_0
  + \boldsymbol{\gamma}_{\text{fam}}^{\top}\,
    \mathbf{1}\{\text{model\_family}_i\}
  + \boldsymbol{\gamma}_{\text{cls}}^{\top}\,
    \mathbf{1}\{\text{classification}_i\}
  + \gamma_{\ell}\,\log\!\bigl(\text{response\_character\_count}_i\bigr),
  \tag{3}\label{eq:logit} \\[10pt]
\log \lambda_i &=
  \beta_0
  + \boldsymbol{\beta}_{\text{fam}}^{\top}\,
    \mathbf{1}\{\text{model\_family}_i\}
  + \boldsymbol{\beta}_{\text{cls}}^{\top}\,
    \mathbf{1}\{\text{classification}_i\} \notag \\[-2pt]
&\quad
  + \beta_{\text{sr}}\,
    \text{unique\_search\_results\_count}_i
  + \beta_{\ell}\,
    \log\!\bigl(\text{response\_character\_count}_i\bigr) \notag \\[-2pt]
&\quad
  + \Bigl(\mathbf{1}\{\text{model\_family}_i\}\!\otimes\!
          \mathbf{1}\{\text{classification}_i\}\Bigr)^{\!\top}\!
    \boldsymbol{\beta}_{\text{fam}\times\text{cls}} \notag \\[-2pt]
&\quad
  + \Bigl(\mathbf{1}\{\text{model\_family}_i\}\!\otimes\!
          \text{unique\_search\_results\_count}_i\Bigr)^{\!\top}\!
    \boldsymbol{\beta}_{\text{fam}\times\text{sr}}
  \tag{4}\label{eq:logmean}
\end{align}

Equation \ref{eq:logit} is the logit probability, where \(\exp(\boldsymbol{\gamma}_{\text{fam}})\) represents the odds-ratios for producing \emph{any} attribution gap by model family, \(\exp(\boldsymbol{\gamma}_{\text{cls}})\) represents the odds-ratios by classification topic, and \(\exp(\gamma_{\ell})\) is the odds-ratio for answer length. The equation contains only main effects (no interactions): the odds of \emph{any} gap depend separately on model family, topic, and answer length.\footnote{For example, \(\exp(\gamma_{\ell})=1.4\) means a one-unit increase in log(response length) multiplies the odds of a gap by 1.4.}

Equation \ref{eq:logmean} is the negative binomial count component whereby \(\exp(\boldsymbol{\beta}_{\text{fam}})\) represents the multiplicative changes in the expected number of uncited website visits by model family, \emph{given that a gap exists}. \(\exp(\boldsymbol{\beta}_{\text{cls}})\) represents the effects by classification topic, and \(\exp(\beta_{\text{sr}})\) is the multiplicative effect of additional search results. A value of 1.30 implies a 30\% larger gap, for example. Equation \ref{eq:logmean} adds \textbf{interaction terms} $\boldsymbol{\beta}_{\text{fam}\times\text{cls}}$, so the effect of a model family can differ by query topic once a gap exists, and $\boldsymbol{\beta}_{\text{fam}\times\text{sr}}$, to account for the varying impact of searching (number of URLs) on the attribution gap by model family.

\subsection{Head-to-Head Model Comparison Regression}
To isolate the number of URLs \emph{each} LLM cites from one additional website search visit, while controlling for question types, we run a second additional regression model. This exploits the LMArena head-to-head design, whereby every question is answered by exactly two model systems. This means that any unobservable query-specific factors driving citation behavior cancel out. This design also ensures that \(\beta_{1m}\) is not biased by systematically ``easy'' or ``hard'' opponents. We collapse each model pair that answers the same query into a single observation \emph{(focal model $-$ opponent)} and run a separate OLS regression for every focal model \(m\), where $i$ now runs from $1$ to $n=6,951$, and $m$ contains 11 focal models. 

\begin{equation}
d_{im}
  = \beta_{0m}
  + \beta_{1m}\,\Delta s_{im}
  + \beta_{2m}\,\Delta \ell_{im}
  + \sum_{k}\gamma^{(m)}_{k}\,D_{ik}
  + \sum_{j}\delta^{(m)}_{j}\,O_{ij}
  + \varepsilon_{im},
\tag{6} \label{eq:pairwise-reg}
\end{equation}

In Equation \ref{eq:pairwise-reg}, $d_{im}$ denotes the \textbf{citation-gap advantage} -- the difference in unique citations produced by the focal model compared to its `opponent', $\text{citations}_{m}-\text{citations}_{\text{opp}}$. The term $\Delta s_{im}$ captures the \textbf{search retrieval difference}, defined as the gap in unique URLs visited by the two systems in its logs, while $\Delta \ell_{im}$ measures the length difference in characters between their answers. The vector $D_{ik}$ comprises topic dummies controlling for the classification of question $i$ (reference category: ``Current affairs \& factual''), and $O_{ij}$ contains dummies identifying which rival model the focal system faces (baseline = the first alphabetic opponent).\footnote{Coefficient interpretation is as follows.  The slope \(\beta_{1m}\) measures \emph{citation efficiency}: it is the expected \emph{additional} citations that model \(m\) produces per \emph{additional} URL it opens, conditional on topic, verbosity, and opponent.  Thus \(\beta_{1m}=0.40\) implies that ten extra retrievals yield roughly four extra citations.  The coefficient \(\beta_{2m}\) captures the effect of answer length (characters) net of retrieval; a significant value would indicate that verbosity itself explains some of the citation edge. Topic dummies enter through \(\gamma^{(m)}_{k}\); positive values mean model \(m\) out-cites its rival in domain \(k\), whereas negative values signal a systematic shortfall.  Opponent-specific effects are absorbed by \(\delta^{(m)}_{j}\), showing how the citation gap changes when \(m\) faces rival \(j\) instead of the baseline opponent.  Finally, \(\beta_{0m}\) gives the baseline citation advantage when the question falls in the reference topic, the opponent is the baseline rival, and retrieval/length differences are zero.}

Each model's \(\beta_{1m}\) is our key quantity of interest and summarizes model \(m\)'s retrieval-to-citation efficiency or yield: \textit{the impact of a marginal website visit on citations}.




\section{Regression Findings}\label{sec:findings}

\subsection{Attribution Gap Regression Model}

\begin{figure}[H]
  \begin{center}
    \vspace{3mm}
    \caption{\centering {\large Expected Attribution Gaps, Predicted (by Model Family)}}
    \vspace{-2mm}
    \centering
    \hspace*{-1cm}
    \includegraphics[scale=0.7]{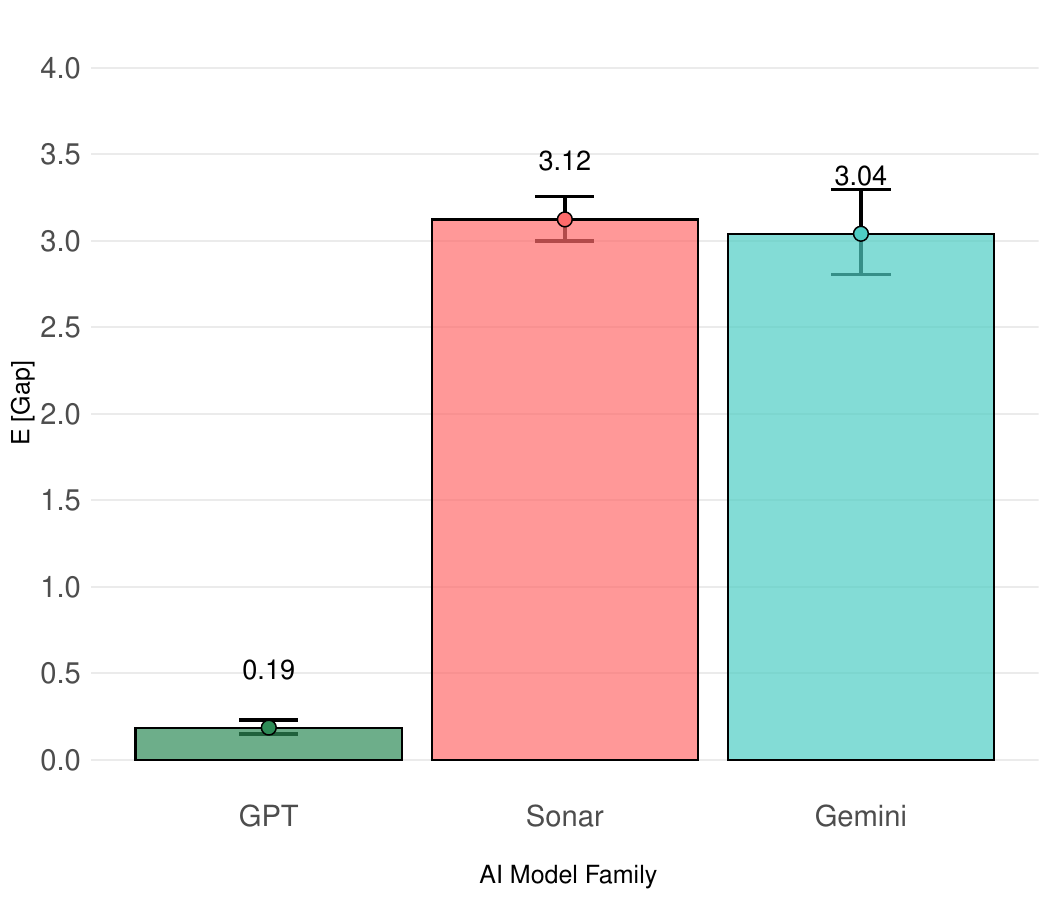}
    \vspace{2mm}
    \caption*{%
      \small Note: Predicted values for number of citations missing relative to web pages consumed. Based on negative binomial hurdle model regression coefficients. Bars show the model's expected citation gap (websites visited in the logs minus websites cited in the output), estimated at the median conversation length and median website visits, without interaction effects included. Showing 95\% confidence intervals calculated with the \textit{emmeans} package in $\mathbb{R}$.}
    \label{fig:citation_gap}
  \end{center}
  \vspace{-10mm}
\end{figure}

The regression results, transformed into probabilities, show notable differences in attribution reliability across AI model families (Table~\ref{tab:expected_gaps_bootstrap} and Figure~\ref{fig:citation_gap}).\footnote{The hurdle model produces raw odds-ratios that compare each model family to the reference category (Gemini). However, for policy interpretation, we need actual probabilities and expected values. To obtain these, we used the fitted model to make predictions for a standardized query: covering Current Affairs \& Factual Information topic classification with median characteristics (5 search results visited, 2,089 character responses). The model's `hurdle' component estimates the probability of having any attribution gap (missing citations \& licensing), while the count component estimates the expected gap size when gaps occur. Multiplying these components gives us the unconditional expectation -- the average number of missing citations per query for each model family.} The `expected attribution gap' is the total prediction from our model: $E[Attribution \ \ Gap] = P(Gap > 0) \ \times \ E[Gap \ | \ Gap > 0]$; whereas the first row (`Probability of a Gap') is only the first term of this equation, $P(Gap > 0)$.

\begin{table}[H]
\centering
\rowcolors{2}{gray!15}{white}
\caption{\centering {\large{Expected Total Attribution Gap by Model Family}}}
\vspace{-2mm}
\begin{tabular}{lrrr}
  \toprule
  & GPT & Gemini & Sonar \\ 
  \midrule
  Probability of a Gap (\%)     & 10.5    & 70.3    & 99.3    \\ 
  Expected Attribution Gap (Total)          & 0.18    & 3.04    & 3.12    \\ 
  95\% Confidence Interval        & [0.15, 0.23] & [2.80, 3.28] & [2.99, 3.25] \\ 
  Standard Error (SE)                  & 0.02   & 0.12   & 0.06   \\ 
  \bottomrule
\end{tabular}
\vspace{2mm}
\caption*{Note: Results from negative binomial hurdle model for Current Affairs \& Factual Information queries with median characteristics (5 search results, 2,089 characters, per query), with bootstrapped mean. Confidence intervals from parametric bootstrap (n=1,000) for total expected attribution gap. Attribution gap is the missing web page (URL) citations per search query relative to relevant websites consumed, measured as a web page gap. This pattern holds consistently across all topic classifications (see Appendix, Table \ref{tab:expected_gaps_all_classifications}).} 
\vspace{-6mm}
\label{tab:expected_gaps_bootstrap}
\end{table}

The GPT-4o models appear to show less exploitative attribution behavior but this is most likely due to them being more circumspect in their disclosures by not showing the full extent of their logs.\footnote{Using OpenAI search in the user interface (UI) shows far more website visits from limited testing.} OpenAI's models have only a 10.5\% probability of having any citation gaps and an expected 0.18 missing citations (relative to relevant websites consumed) per query (Table \ref{tab:expected_gaps_bootstrap}). Sonar exhibits citation gaps in 99.3\% of queries with 3.12 expected missing citations per query in total. Gemini falls somewhere between these extremes with citation gaps in 70.3\% of queries and 3.04 expected missing citations in total per query. 

Sonar's near-universal citation gap (99.3\% of queries) makes it particularly concerning for applications requiring source transparency. Gemini is arguably relying too heavily on internal knowledge given that a large portion of its zero attribution gap is due to it having zero (disclosed) website searches, at 34\% of queries.\\[-4mm]

Lastly, in terms of other factors driving \emph{how large} the gap is once it appears (the count component), we can look at the incidence-rate ratio (IRR).\footnote{In count-data models (Poisson, negative binomial, zero-inflated, hurdle, etc.) the regression is fitted on a log scale. Exponentiating a coefficient converts it to an incidence-rate ratio (IRR). The IRR tells you the multiplicative change in the expected count when the predictor increases by one unit, holding everything else constant.} Longer answers actually \emph{shrink} the gap size: When logged response character count doubles (from e.g., 100 to 200 characters), the expected number of missing citations decreases by about 11\% (IRR 0.89 or so). Reading more web pages from search inflates the gap: every extra page the LLM system reads raises the expected uncited count by 13\% (IRR $\approx$ 1.13).

\subsection{Head-to-Head Model Results: Citation efficiency}
In this section, we focus on $\beta_{1m}$ coefficient from our head-to-head regressions (Equation \ref{eq:pairwise-reg}). This coefficient tells us how many more citations the focal model generates for each additional search result compared to its opponent. A smaller coefficient signals poorer performance: the focal model converts each extra relevant page it opens into fewer citations than its comparator does.
For example, if Sonar's coefficient is 0.4, this means that when it undertakes five additional relevant web page visits than GPT-4o on a question it is expected to have $5 \times 0.4 = 2$ more citations than GPT-4o. 264 coefficients are estimated in total, running a separate regression for each of the 11 models (24 parameters estimated for each model).\footnote{1 $\times$ Intercept; 11 $\times$ topic dummies (classifications), 1 $\times$ focal-search-diff (number of web pages), 10 $\times$ opponent-model dummies, 1 $\times$ focal-length-diff.} 

\begin{figure}[H]
\begin{center}
\vspace{3mm}
        \caption{\centering {\large{Focal Model: Citation difference per extra URL visited}}}
        \vspace{-1mm}
        \centering
        \hspace*{-1cm}
        \includegraphics[scale=.70]{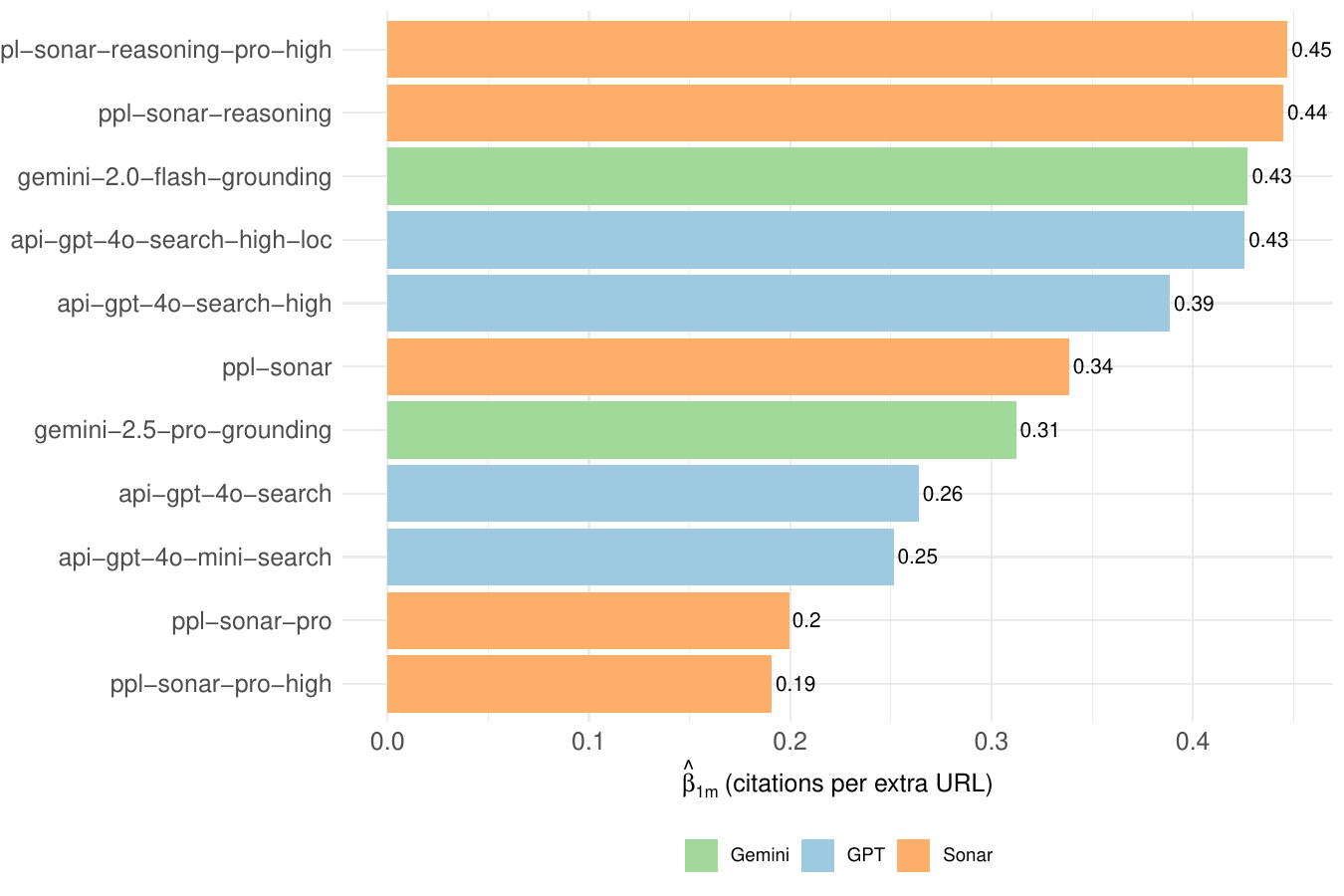}
        \vspace{2mm}
        \caption*{Note: Extra citations gained for each additional URL the focal model opens (differences between models). This holds match-up effects constant, isolating technology effects. Regression coefficient $\beta_{1m}$ shown, predicting differences in citations for model pairs, for a given query. See equation above.}
        \label{fig:reg2}
        \end{center}
        \vspace{-8mm}
\end{figure}

We plot $\beta_{1m}$ in Figure \ref{fig:reg2} below (Appendix Table \ref{tab:appendix_beta1_compact}). Almost all coefficients from the regression are positive and highly significant.\footnote{Among the 264 coefficients estimated, \texttt{focal\_search\_diff} is significant (and positive) for every model -- extra retrieval still translates into more citations, although the payoff varies. Answer length matters (positive) for the GPT-mini, GPT-search, Sonar-pro, Sonar, and Sonar-reasoning families, but not for the ``high'' search variants or Gemini models. One notable exception: \texttt{ppl-sonar-reasoning-pro-high} shows a significant \emph{negative} length effect ($-1.35 \times 10^{-4}$, $p = 6.75 \times 10^{-3}$), suggesting longer answers actually hurt citation performance for this model. Topic effects are sparse and model-specific. Mental \& Physical Health lowers citation edge for GPT-4o variants but raises it for basic Sonar-pro. Finance boosts citation edge for Sonar-pro-high but suppresses it for Sonar-reasoning and Gemini-2.5. Opponent dummies are often significant and large.}

Figure \ref{fig:reg2} below shows that best-in-class variants (Sonar-reasoning-pro-high, Gemini-flash-grounding, GPT-4o-search-high-loc) yield \(\sim0.43\) citations per extra URL; whereas baseline variants (ppl-sonar-pro-high) return only \(\sim0.19\). This implies that the best model converts every extra retrieved URL into $\approx$ 0.45 additional citations (compared to its competitor models), whereas the weakest model variant converts that same extra URL into $\approx$ 0.19 citations. The span is therefore about $0.45 - 0.19 \approx 0.26$ citations per URL, showing that RAG implementation choices can more than double the payoff from each additional page the model visits. This illustrates just how wide the performance window is, and thus how much room developers (or regulators) have to raise low performers up to the current best practice. 

More specifically, our results show that RAG implementation type (technology) is more important than model family (e.g., OpenAI vs. Gemini): 
\begin{itemize}
\item An ANOVA test shows that within-family variation (the variance) of search coefficients $\beta_{1m}$ is almost 8 times larger than between-family variation. This means that differences in model behavior is far greater \textit{within} model families than between them.\footnote{We ran a one–way ANOVA on the eleven citation–efficiency coefficients ($\widehat{\beta}_{1m}$ on \texttt{focal\_search\_diff}) grouped by provider family. The \emph{within–family} mean square is nearly eight times the \emph{between–family} mean square:

\[
\mathrm{MS}_{\text{within}} \;=\; 0.0116,
\qquad
\mathrm{MS}_{\text{between}} \;=\; 0.0015,
\qquad
F_{(2,\,8)} \;=\; \frac{0.0015}{0.0116} \;\approx\; 0.13,
\; p = 0.88.
\]

Thus, variability inside each provider family dominates the small
differences between family means: implementation choices explain
\(\sim 8\times\) more of the spread in citation efficiency than the family label itself.}

\item Within Sonar alone, upgrading to the \emph{reasoning} tier more than doubles efficiency (from 0.19-0.20 $\to$ 0.44-0.45), with Sonar-reasoning (0.44) vs Sonar-pro-high (0.19) = 0.25 difference. For GPT models: GPT-4o-search-high-loc (0.42) vs GPT-4o-mini (0.25) = 0.17 difference.

\item \textit{Location signals} matter too. Adding a country code to GPT-4o raises search-citation coefficient efficiency by roughly 10\,\% (0.39 $\to$ 0.43), confirming that retrieval relevance translates directly into better attribution (though it is unclear why this is the case). Practically, this effect size is moderate since for a model getting 10 extra search results, location signals would generate $\approx$ 0.37 additional citations. But we know that local search operates differently, both in traditional search and increasingly with LLMs \citep{rollison2025chatgptLocal}.
\end{itemize}

Lastly, the regression results show that more basic models (GPT-mini, GPT-search, and basic Sonar models) compensate for lower search efficiency through verbosity -- they need longer answers to achieve similar citation performance. Advanced (`high' variant and Gemini) models achieve higher citation rates through superior search utilization, making verbosity unnecessary to achieve improved citation rates.

\section{Policy Implications} \label{sec:discussion}

Without standardized telemetry — comprehensive logs and traces of what an LLM retrieves and cites — no transparent and competitive market for
licensing, revenue-sharing, or other content-monetization schemes can easily emerge. Publishers need hard numbers on how often their pages power an answer in order to automate royalty or revenue-share flows; regulators need the same auditable data to enforce forthcoming disclosure rules in jurisdictions such as the European Union (EU) and California. In short, richer telemetry is \textit{the} prerequisite for both commercial remuneration and public oversight.

The technical pieces already exist for full disclosure of an LLM's search and citation trace; what remains is coordination. The key challenge is to persuade providers to adopt a common telemetry standard — and to ensure buyers and regulators can incentivize its provision.

Modern observability frameworks — LangSmith, Langfuse, Phoenix, and the GenAI semantic-conventions in OpenTelemetry — allow for recording an end-to-end \emph{search trace}, which can detail the search activities that an LLM RAG system undertakes when trying to find the most relevant context for the user. One way to think of traces is as a collection of structured logs with context, correlation, and hierarchy baked in. Each web page retrieval, reranking, and generation step by the LLM can be logged as an OpenTelemetry span, provided every document is tagged with a stable \emph{Source ID}, typically a hash of the original URL or file.\footnote{With a span representing a unit of work or operation and are the building blocks of traces. See: \url{https://opentelemetry.io/docs/concepts/signals/traces/}.} 

Because hashes are one-way fingerprints, they enable later verification without revealing copyrighted text. The same span can carry the numerical relevance score (`how relevant was this webpage to the LLM's answer') produced by the vector store, BM25 index, or cross-encoder, as long as that score is preserved in the trace \citep{langchain_rag_citations, ryaboy2025crossencoders}. If each retrieved document carries two stable fields — say
\verb|llm.retrieval.ids| (a hash or URL that uniquely identifies the page) and \verb|llm.retrieval.scores| (its relevance score or rank) — \emph{and} those fields are propagated from the retrieval step all the way to the final API response, then anyone inspecting the trace can, in theory:

\begin{enumerate}[nosep]
  \item Enumerate every page the model actually saw;
  \item Check which of those pages were later cited in the answer; and
  \item Compare the relevance scores of cited pages with those that were ignored.
\end{enumerate}

In short, the full provenance of ``pages viewed'' versus ``pages credited'' becomes auditable.\footnote{\url{https://opentelemetry.io/docs/specs/semconv/gen-ai/}}

OpenTelemetry-like open protocols can be a lightweight standard that enables comparison and validation of LLM behavior across models. And adopting telemetry protocols (standards) requires only incremental changes to today's open-source stacks.\footnote{In LangChain, a single line of code drops the hash and score into \texttt{Document.metadata}. LangSmith then records them automatically in each trace span \citep{langchain_rag_citations}. Langfuse performs the same mapping when it converts LangChain calls into OpenTelemetry. Phoenix ingests any span that follows the GenAI conventions, meaning dashboards that plot ``high-score pages not cited'' or ``low-score pages that slipped through'' can be deployed without further engineering.}

The real challenge, perhaps, lies in constructing the market and regulatory incentives to advance widespread adoption and disclosure. Transparent traces unlock clear, quantifiable monetary benefits for providers. API buyers in legal, medical, and financial sectors increasingly require provenance guarantees. This means that model developers that expose richer evidence trails could, therefore, command premium pricing and benefit from greater demand in these compliance-sensitive sectors. However, the model developers may be reticent to undertake such additional disclosures without liability safeguards or more reliable RAG pipelines.

Yet our findings also point toward possible paths forward. A business model for LLM search could create a mutually beneficial exchange of content for web traffic or direct purchases. Commercial queries present a fairly distinctive pattern in our analysis. Shopping and commercial-intent queries show attribution gaps 76\% larger than other categories, yet these domains offer opportunities for such mutually beneficial arrangements. Early web advertising models, where attribution facilitated click-through and conversion, provide relevant precedents for sustainable approaches.


\section{Conclusion}\label{sec:conclusion}
This study provides one of the first systematic empirical audits of attribution practices in commercial search-augmented LLMs. When LLMs exploit content from platforms without proper attribution, they undermine the economic incentives that sustain high-quality information production. Returning to our opening quote by the late economist Joan Robinson -- on the notion that exploitation under capitalism involves the absence of commercial relations as well as its presence -- we find similar twin forces at work in our analysis. Gemini systematically excludes the world wide web's content ecosystem when answering questions as its form of exploitation, while Perplexity exploits through the opposite behavior, overly zealous consumption of web content without commensurate attributions.

We find a substantial ``attribution gap'' between relevant content consumed from websites and attribution practices that threatens the sustainability of the digital content ecosystem. Our analysis of $\approx$14,000 real-world interactions demonstrates that leading AI systems systematically consume web content without adequate attribution, with Gemini providing no citations in 92\% of its responses and Perplexity visiting approximately a dozen relevant websites while crediting only a few. 

A negative binomial hurdle model shows that the average query answered by Gemini or Sonar leaves about 3 relevant websites uncited, whereas GPT-4o’s tiny uncited gap is best explained by its selective log disclosures rather than by better attribution. GPT-4o models remain difficult to audit due to what appears to be stricter pre-filtering of its model search logs.

The dramatic variation in citation efficiency between models when answering the same question -- from 0.19 to 0.45 citations per additional web page visited -- illustrate that attribution gaps result from design choices, not simply technical limitations. The best-performing systems show that transparent, comprehensive attribution is technically feasible today (even if accurate citations and output remain a real ongoing issue).


Closing the attribution gap is, therefore, less a technical hurdle than one of proper coordination and market incentives. Standardizing two telemetry fields — document hashes and relevance scores — would allow anyone to verify which information an LLM consumed and how faithfully it credited that information. Observability tools already provide the necessary plumbing. What remains is a collective decision by model providers to enable it and by developers and buyers to reward those who do. Transparent search traces would strengthen incentives for high-quality content and give users a robust evidentiary basis for trusting machine-generated answers.

Our results demonstrate that transparency in the web sources consumed and cited by LLMs when answering user queries is fundamentally an engineering choice.\footnote{This aligns with emerging research on LLM citation evaluation frameworks \citep{gao2023enabling} and attribution methods in scientific literature \citep{saxena2024attribution, najjar2025leveraging}, which show that technical solutions are available but underutilized. Source-aware training and fine-tuning have also been shown to improve citation behavior \citep{khalifa2024source}. See also \citet{asai2023self} and \citet{borgeaud2022improving}.}

\newpage
\bibliography{main}

\cleardoublepage
\section{Appendix} \label{sec:appendix}

\subsection{Full Hurdle-Model Specification and Diagnostics} \label{app:hurdle}

The regression is estimated with R's \texttt{pscl::hurdle()} function, which maximizes a likelihood that combines a \emph{binary gate} for perfect attribution and a \emph{zero-truncated} negative binomial (NB) for positive gaps.  Writing $\pi_i\equiv\Pr(Y_i=0)$, the log-likelihood is  

\[
\ell(\boldsymbol{\gamma},\boldsymbol{\beta},\alpha)=
\sum_{i:\,y_i=0}\!\log\pi_i
\;+\;
\sum_{i:\,y_i>0}\!\Bigl\{
\log(1-\pi_i)
+\log f_{\mathrm{NB}}(y_i\mid\lambda_i,\alpha)
-\log\bigl[1-f_{\mathrm{NB}}(0\mid\lambda_i,\alpha)\bigr]
\Bigr\},
\tag{A1}
\]

with
\(\boldsymbol{\gamma}=(\gamma_0,\gamma_1,\gamma_2,\gamma_3)^{\!\top}\)
for the gate and
\(\boldsymbol{\beta}=(\beta_0,\beta_1,\beta_2,\beta_3,\beta_4)^{\!\top}\)
for the count component.  
The NB kernel is parameterised by its mean \(\lambda_i\) and
over-dispersion \(\alpha>0\):

\[
f_{\mathrm{NB}}(y\mid\lambda_i,\alpha)=
\frac{\Gamma\!\bigl(y+\alpha^{-1}\bigr)}
     {\Gamma(\alpha^{-1})\,y!}\,
\bigl(\alpha\lambda_i\bigr)^{y}\,
\bigl(1+\alpha\lambda_i\bigr)^{-\bigl(y+\alpha^{-1}\bigr)},
\qquad
\text{Var}(Y_i\mid Y_i\ge1)=\lambda_i+\alpha\lambda_i^{2}.
\tag{A2}
\]

Likelihood ratio tests can assess whether the count process is better modeled as Poisson or negative binomial (null hypothesis $H_0: \alpha = 0$), and whether inclusion of the interaction term improves model fit. A negative binomial outperforms. Model fits (information criteria) are superior with interaction terms and shows domains where the effect of a model family changes the expected gap magnitude. 

With interaction terms the zero-inflated model performs better in a Vuong test than a hurdle model. Our hurdle model though aligns better with theoretical understanding of citation process. We believe interpretability advantages outweigh improvements though. BFGS optimisation converged in 54 iterations, giving $\widehat{\theta}=5.69$. 

\begin{figure}[H]
\begin{center}
\vspace{3mm}
        \caption{\centering {\large{Attribution Gap Histogram vs Negative Binomial (Red)}}}
        \vspace{-2mm}
        \centering
        \hspace*{-1cm}
        \includegraphics[scale=.70]{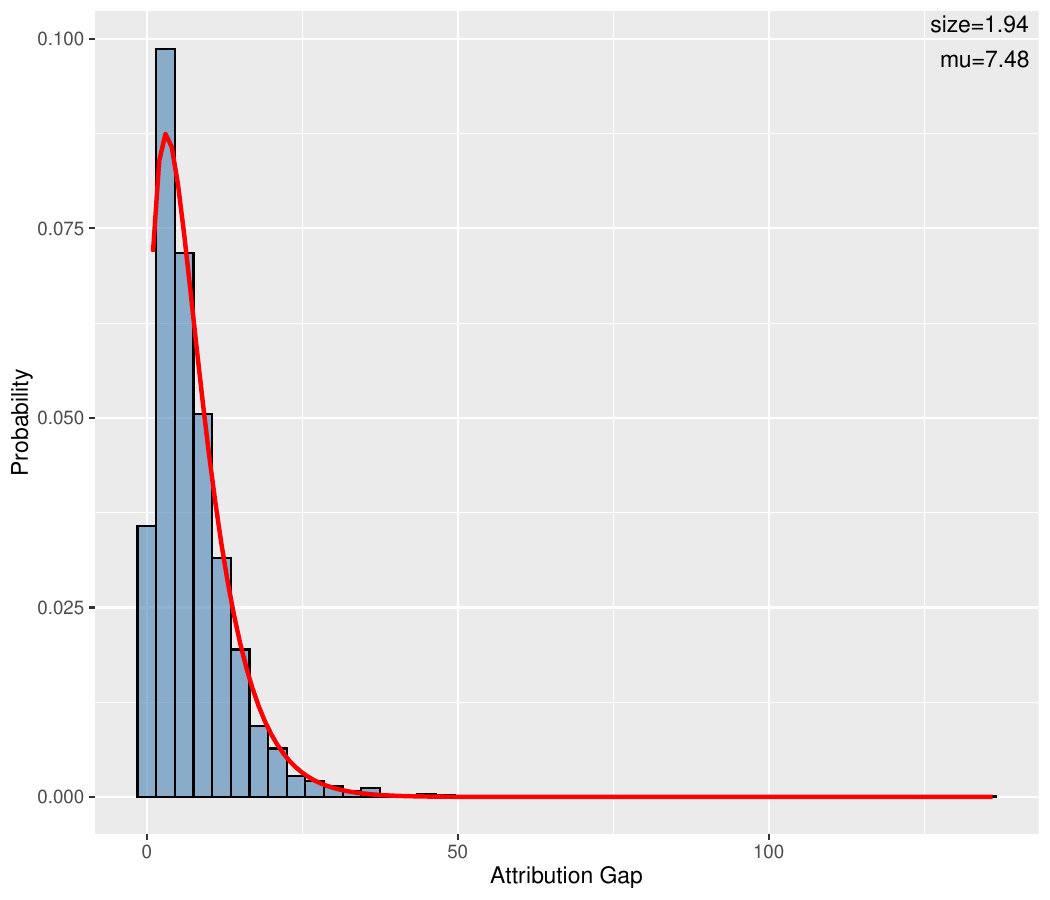}
        \vspace{-1mm}
        \caption*{Note: Attribution gap (relevant sites visited in logs minus websites cited by in-text URLS) shows a negative binomial distribution (red over-plot).}
        \label{fig:negative_binomial}
        \end{center}
        \vspace{-1mm}
\end{figure}

\subsection{Detailed Model Results}

\begin{figure}[H]
  \begin{center}
    \vspace{3mm}
    \caption{\centering {\large Expected Citation Gaps by Regression Model (Query Classification and Model Family)}}
    \vspace{-5mm}
    \centering
    \hspace*{-1cm}
    \includegraphics[scale=0.70]{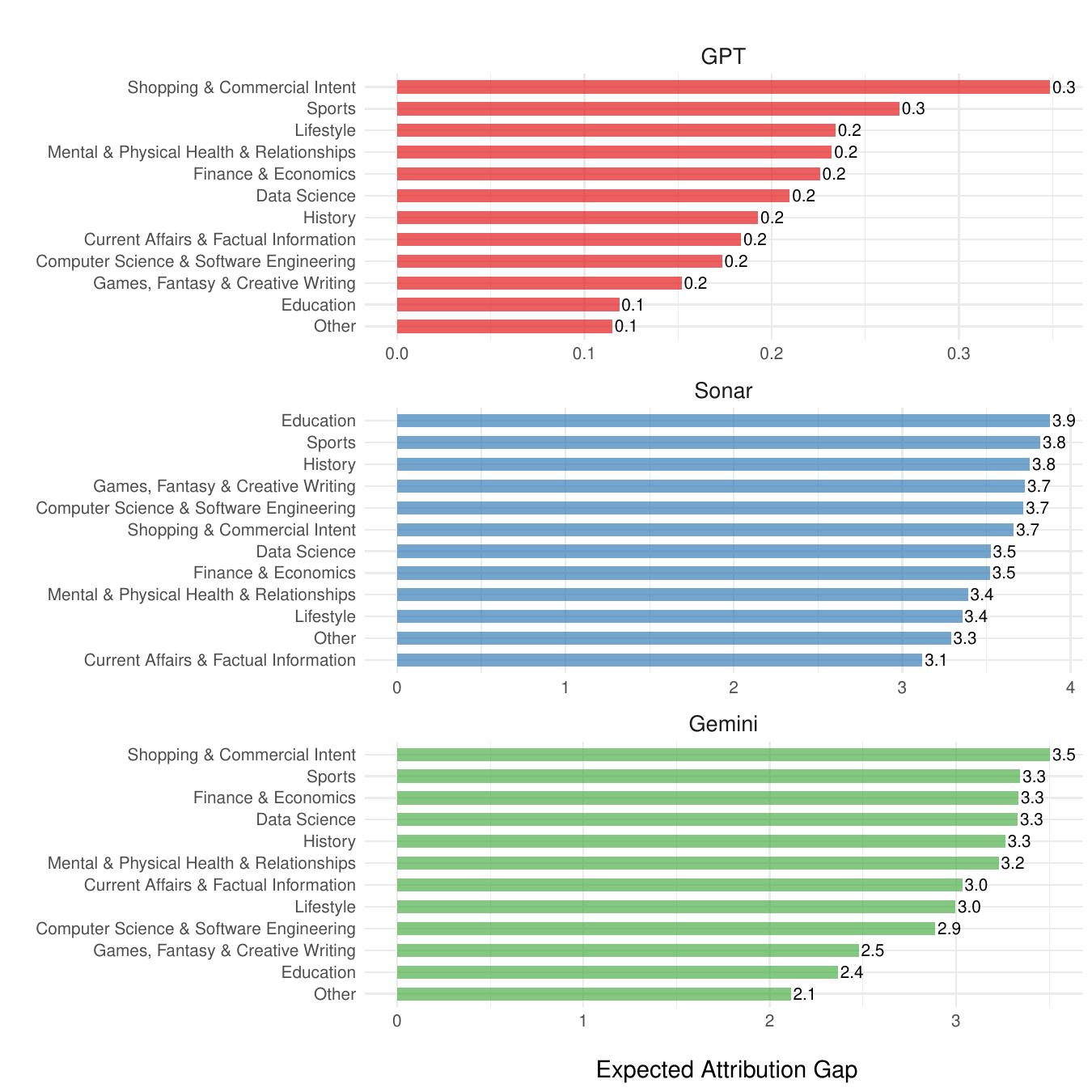}
    \vspace{3mm}
    \caption*{%
      \small Note: Bars show the expected citation gap (relevant websites visited in the logs minus websites cited in the LLM's output) for each query classification. Classification labels are ordered by gap magnitude within each panel. See Table \ref{tab:expected_gaps_bootstrap} above for method.}
    \label{fig:results_classification}
  \end{center}
  \vspace{-10mm}
\end{figure}

Figure \ref{fig:results_classification} shows topic specific gaps, as predicted by our model. `Shopping \& Commercial Intent', perhaps surprisingly, shows the largest gap across models -- being 0.2 lower for the Sonar model family. This could reflect a greater selectivity in which results are shown by the model given potential greater commercial incentives to `get it right', or to abide by pre-existing licensing agreements. But it also could simply highlight the absence of a worked out business model in this area, despite large opportunities to monetize website traffic in this area of LLM search.\\

\begin{table}[H]
\centering
\caption{\large{Count Model Coefficients (No Interactions)}} 
\label{tab:count_main_effects}
\begin{tabular}{lrrrr}
  \hline
Term & Estimate & Std. Error & z value & p-value \\ 
  \hline
(Intercept) & 2.163 & 0.078 & 27.66 & 0.000 \\ 
  modelfamilyGPT & -1.099 & 0.156 & -7.05 & 0.000 \\ 
  modelfamilySonar & 0.207 & 0.054 & 3.81 & 0.000 \\ 
  classificationCurrent Affairs \& Factual Information & -0.021 & 0.053 & -0.39 & 0.695 \\ 
  classificationData Science & 0.057 & 0.064 & 0.89 & 0.371 \\ 
  classificationEducation & -0.115 & 0.091 & -1.26 & 0.209 \\ 
  classificationFinance \& Economics & 0.034 & 0.068 & 0.50 & 0.615 \\ 
  classificationGames, Fantasy \& Creative Writing & -0.053 & 0.074 & -0.72 & 0.473 \\ 
  classificationHistory & -0.008 & 0.087 & -0.09 & 0.929 \\ 
  classificationLifestyle & -0.028 & 0.064 & -0.43 & 0.665 \\ 
  classificationMental \& Physical Health \& Relationships & 0.050 & 0.077 & 0.66 & 0.512 \\ 
  classificationOther & -0.093 & 0.065 & -1.43 & 0.154 \\ 
  classificationShopping \& Commercial Intent & 0.021 & 0.061 & 0.35 & 0.727 \\ 
  classificationSports & -0.033 & 0.091 & -0.37 & 0.712 \\ 
  search results count & 0.125 & 0.003 & 39.74 & 0.000 \\ 
  log(response length) & -0.171 & 0.008 & -20.24 & 0.000 \\ 
  $\log(\theta)$ & 1.740 & 0.031 & 55.39 & 0.000 \\ 
   \hline
\end{tabular}
\begin{minipage}{\linewidth}
\vspace{2mm}
\footnotesize Note: Gemini as baseline model. Showing coefficients (main effects) from the count component of the negative binomial hurdle model.
\end{minipage}
\end{table}

\begin{table}[H]
\centering
\caption{\large{Hurdle Model Coefficients (No Interactions)}} 
\label{tab:zero_main_effects}
\begin{tabular}{lrrrr}
  \hline
Term & Estimate & Std. Error & z value & p-value \\ 
 \hline
(Intercept) & -0.729 & 0.207 & -3.52 & 0.000 \\ 
  modelfamilyGPT & -3.140 & 0.069 & -45.66 & 0.000 \\ 
  modelfamilySonar & 1.601 & 0.061 & 26.28 & 0.000 \\ 
  classificationCurrent Affairs \& Factual Information & 0.195 & 0.096 & 2.03 & 0.043 \\ 
  classificationData Science & 0.268 & 0.117 & 2.29 & 0.022 \\ 
  classificationEducation & -0.245 & 0.144 & -1.69 & 0.090 \\ 
  classificationFinance \& Economics & 0.341 & 0.121 & 2.83 & 0.005 \\ 
  classificationGames, Fantasy \& Creative Writing & -0.265 & 0.108 & -2.45 & 0.014 \\ 
  classificationHistory & 0.398 & 0.169 & 2.36 & 0.018 \\ 
  classificationLifestyle & 0.177 & 0.111 & 1.59 & 0.113 \\ 
  classificationMental \& Physical Health \& Relationships & 0.189 & 0.144 & 1.31 & 0.190 \\ 
  classificationOther & -0.529 & 0.098 & -5.39 & 0.000 \\ 
  classificationShopping \& Commercial Intent & 0.567 & 0.115 & 4.93 & 0.000 \\ 
  classificationSports & 0.577 & 0.151 & 3.81 & 0.000 \\ 
  log(response length) & 0.165 & 0.024 & 6.92 & 0.000 \\ 
   \hline
\end{tabular}
\begin{minipage}{\linewidth}
\vspace{2mm}
\footnotesize Note: Gemini as baseline model. Showing coefficients (main effects) from the hurdle component of the negative binomial hurdle model.
\end{minipage}
\end{table}

\begin{table}[H]
\centering
\small
\rowcolors{2}{gray!15}{white}
\caption{\centering {\large{Expected Citation Gaps by Model Family and Topic Classification}}}
\begin{tabular}{l@{\hspace{8pt}}rr@{\hspace{12pt}}rr@{\hspace{12pt}}rr}
\toprule
& \multicolumn{2}{c}{GPT-4o} & \multicolumn{2}{c}{Gemini} & \multicolumn{2}{c}{Sonar} \\
\cmidrule(lr){2-3} \cmidrule(lr){4-5} \cmidrule(lr){6-7}
Classification & Est. & 95\% CI & Est. & 95\% CI & Est. & 95\% CI \\
\midrule
Computer Science \& Software Engineering & 0.17 & (0.14--0.22) & 2.89 & (2.63--3.14) & 3.72 & (3.55--3.89) \\
Current Affairs \& Factual Information & 0.18 & (0.15--0.23) & 3.03 & (2.78--3.29) & 3.12 & (3.00--3.26) \\
Data Science & 0.21 & (0.16--0.27) & 3.33 & (2.97--3.73) & 3.52 & (3.34--3.74) \\
Education & 0.12 & (0.08--0.19) & 2.37 & (1.97--2.78) & 3.88 & (3.57--4.19) \\
Finance \& Economics & 0.23 & (0.18--0.29) & 3.34 & (2.95--3.73) & 3.52 & (3.34--3.72) \\
Games, Fantasy \& Creative Writing & 0.15 & (0.11--0.21) & 2.48 & (2.18--2.82) & 3.73 & (3.54--3.92) \\
History & 0.19 & (0.13--0.30) & 3.27 & (2.77--3.80) & 3.76 & (3.45--4.10) \\
Lifestyle & 0.23 & (0.17--0.31) & 3.00 & (2.69--3.34) & 3.36 & (3.19--3.54) \\
Mental \& Physical Health \& Relationships & 0.23 & (0.17--0.32) & 3.23 & (2.76--3.73) & 3.39 & (3.17--3.63) \\
Other & 0.11 & (0.09--0.15) & 2.11 & (1.88--2.39) & 3.29 & (3.13--3.47) \\
Shopping \& Commercial Intent & 0.35 & (0.27--0.45) & 3.51 & (3.16--3.85) & 3.66 & (3.48--3.86) \\
Sports & 0.27 & (0.19--0.38) & 3.35 & (2.85--3.95) & 3.82 & (3.59--4.07) \\
\bottomrule
\end{tabular}
\vspace{2mm}
\caption*{Note: Total expected attribution gap from negative binomial hurdle model. Predictions are at the median query (5 search results and 2,089 characters). Confidence intervals from parametric bootstrap (n=1,000). Est. = Expected citation gaps per query.}
\label{tab:expected_gaps_all_classifications}
\end{table}

\subsection{Head-to-Head Regressions: Table of results}

\newcolumntype{L}[1]{>{\raggedright\arraybackslash}p{#1}}
\begin{table}[htbp]
  \centering
  \caption{\large{Incremental citations per extra URL opened, by model variant}}
  \label{tab:appendix_beta1_compact}
  \begin{tabular}{p{8cm} l r r r r}
    \toprule
    \textbf{Model variant} & \textbf{Family} &
    $\mathbf{n}$ & $\boldsymbol{\hat\beta_{1m}}$ & \textbf{SE} & $\boldsymbol{p}$ \\
    \midrule
    \texttt{ppl-sonar-reasoning-pro-high} & Sonar  &  848 & 0.447 & 0.015 & 1.4e-138 \\
    \texttt{ppl-sonar-reasoning}          & Sonar  & 1635 & 0.445 & 0.012 & 4.3e-215 \\
    \texttt{gemini-2.0-flash-grounding}   & Gemini & 1189 & 0.427 & 0.013 & 6.0e-172 \\
    \texttt{api-gpt-4o-search-high-loc}   & GPT    & 1222 & 0.426 & 0.010 & 6.1e-240 \\
    \texttt{api-gpt-4o-search-high}       & GPT    & 1695 & 0.389 & 0.010 & 1.3e-224 \\
    \texttt{ppl-sonar}                    & Sonar  & 1200 & 0.339 & 0.012 & 1.8e-128 \\
    \texttt{gemini-2.5-pro-grounding}     & Gemini & 1200 & 0.312 & 0.011 & 4.8e-130 \\
    \texttt{api-gpt-4o-search}            & GPT    & 1188 & 0.264 & 0.012 & 2.5e-89  \\
    \texttt{api-gpt-4o-mini-search}       & GPT    & 1158 & 0.251 & 0.012 & 1.1e-80  \\
    \texttt{ppl-sonar-pro}                & Sonar  & 1208 & 0.200 & 0.014 & 9.5e-46  \\
    \texttt{ppl-sonar-pro-high}           & Sonar  & 1359 & 0.191 & 0.012 & 9.9e-51  \\ 
    \bottomrule
  \end{tabular}

  \vspace{1ex}
  \captionsetup{labelformat=empty}
  \caption*{\footnotesize
    Note: $n$ = pair-wise query comparisons involving the focal model. 
    $\hat\beta_{1m}$ is the coefficient on \texttt{focal\_search\_diff}, interpreted as the expected number of additional citations produced for each extra URL the focal model opens.
  }
\end{table}
\end{document}